\documentclass[aps,ams,amsmath,prx,twocolumn,superscriptaddress]{revtex4-1}
\usepackage{graphics}
\usepackage{graphicx}
\usepackage{amsmath}
\usepackage{amsfonts}
\usepackage{bm}
\usepackage{color}
\usepackage{bbm}
\usepackage{hyperref}
\usepackage{soul}
\def\bm#1{\mbox{\boldmath{$#1$}}}

\newcommand{\be}{\begin{equation}}
\newcommand{\ee}{\end{equation}}
\def\boldsymbol#1{\mbox{\boldmath$#1$}}
\hypersetup{colorlinks=true,linkcolor=blue,anchorcolor=blue,citecolor=blue,filecolor=blue,urlcolor=blue,bookmarksnumbered=true,pdfview=FitB}

\newcommand{\q}{\mathbf{q}}

\renewcommand{\r}{\mathbf{r}}

\newcommand{\beq}{\begin{equation}}
\newcommand{\eeq}{\end{equation}}
\newcommand{\beqa}{\begin{eqnarray}}
\newcommand{\eeqa}{\end{eqnarray}}
\newcommand{\bea}{\begin{eqnarray}}
\newcommand{\eea}{\end{eqnarray}}
\usepackage{changes}
\usepackage{ulem}
\usepackage{soul}
\usepackage{amssymb}
\usepackage{array}
\usepackage{amsmath}
\usepackage{graphicx}\usepackage{graphics}
\usepackage{dcolumn}
\usepackage{bm}\usepackage{varioref}

\newcommand{\calE}{\boldsymbol{\mathcal{E}}}

\begin{document}

\title{Hydrodynamic electron transport near charge neutrality}

\author{Songci Li}
\affiliation{Department of Physics, University of Wisconsin-Madison, Madison, Wisconsin 53706, USA}

\author{Alex Levchenko}
\affiliation{Department of Physics, University of Wisconsin-Madison, Madison, Wisconsin 53706, USA}

\author{A. V. Andreev}
\affiliation{Department of Physics, University of Washington, Seattle, Washington 98195, USA}

\affiliation{Skolkovo  Institute of  Science  and  Technology,  Moscow,  143026,  Russia}
\affiliation{ L. D. Landau Institute for Theoretical Physics, Moscow, 119334, Russia}

\date{August 9, 2020}

\begin{abstract}
We develop the theory of hydrodynamic electron transport in a long-range disorder potential for conductors in which the underlying electron liquid lacks Galilean invariance. For weak disorder, we express the transport coefficients of the system in terms of the intrinsic kinetic coefficients of the electron liquid and the correlation function of the disorder potential. We apply these results to analyze the doping and temperature dependence of transport coefficients of graphene devices. We show that at charge neutrality, long-range disorder increases the conductivity of the system above the intrinsic value. The enhancement arises from the predominantly vortical hydrodynamic flow caused by local deviations from charge neutrality. Its magnitude is inversely proportional to the shear viscosity of the electron liquid and scales as the square of the disorder correlation radius. This is qualitatively different from the situation away from charge neutrality. In that case, the flow is predominantly potential, and produces negative viscous contributions to the conductivity, which are proportional to the sum of shear and bulk viscosities, and inversely proportional to the square of disorder correlation radius.
\end{abstract}

\maketitle

\section{Introduction}\label{sec:intro}

There is mounting evidence that under certain conditions hydrodynamic effects predicted by Gurzhi \cite{Gurzhi-UFN} play an important role in electron transport properties of semiconductors \cite{Molenkamp,deJong,Predel,Gao,Manfra}, and monolayer or bilayer graphene-based devices \cite{Bandurin-1,Crossno,Ghahari,Kumar,Morpurgo,Bandurin-2,Berdyugin,Hone} (see recent reviews \cite{Spivak,NGMS-HydroReview,Lucas-Fong-HydroReview} on these topics and references therein). In addition to the transport measurements, several imaging techniques \cite{Jura,Ensslin,Ilani,Yacoby,Jayich} have been implemented to directly map out profiles of electron flow in narrow graphene channels and mesoscopic Ga(Al)As. Furthermore, possible signatures of hydrodynamic electron flow have been identified in transport measurements in quasi-two-dimensional delafossite metals PdCoO$_2$ and PtCoO$_2$ \cite{Moll,Nandi}, and Dirac or type-II Weyl semimetallic conductors \cite{Gooth-PtSn4,Gooth-WP2}. This motivates further studies of hydrodynamic effects in various systems including, for example, one-dimensional electron liquids in quantum wires \cite{LMRM,DeGottardi,Matveev}, and electronic double-layers \cite{Gorbachev,ALA,CAL,Leo}.

One of  the most salient signatures of hydrodynamic effects in electron transport is the violation of Matthiessen's rule, according to which the resistivity should be proportional to the sum of momentum relaxation rates due to various scattering processes. Within this paradigm, momentum-conserving electron-electron (\emph{ee}) collisions should not affect the resistivity.

In the hydrodynamic regime, the resistivity depends on the rate of momentum-conserving \emph{ee} scattering via the viscosity and other dissipative characteristics of the electron liquid. This dependence arises from correlations between \emph{ee} scattering and the underlying disorder and/or confining potential, which are ignored in the derivation of the Matthiessen's rule.

Moreover, as the rate of \emph{ee} scattering increases, the resistance of the system often decreases. This effect, first pointed out by Gurzhi, occurs not only in finite geometries, such as the Poiseuille flow \cite{Gurzhi-Poiseuille,Molenkamp,deJong}, or point contacts \cite{Kumar,Guo-1}, but also in the bulk~\cite{Hruska,AKS,Pal,Lucas,Guo-2,LXA}.

On the other hand, disorder usually increases the resistivity. Indeed, away from charge neutrality, acceleration of the liquid by the external electric field would result in vanishing resistivity in the absence of disorder. In particular, for  Galilean invariant liquids the systems resistivity is proportional to the disorder strength \cite{AKS}. In the more general case~\cite{Hartnoll,Sachdev,Aleiner,Mirlin,Lucas,PDL}, in which the electron liquid does not possess Galilean invariance, the  resistivity of the system away from charge neutrality is still enhanced by disorder, although the dependence of the resistivity on the disorder strength is more complicated.

\begin{figure}[t!]
  \centering
  \includegraphics[width=3.5in]{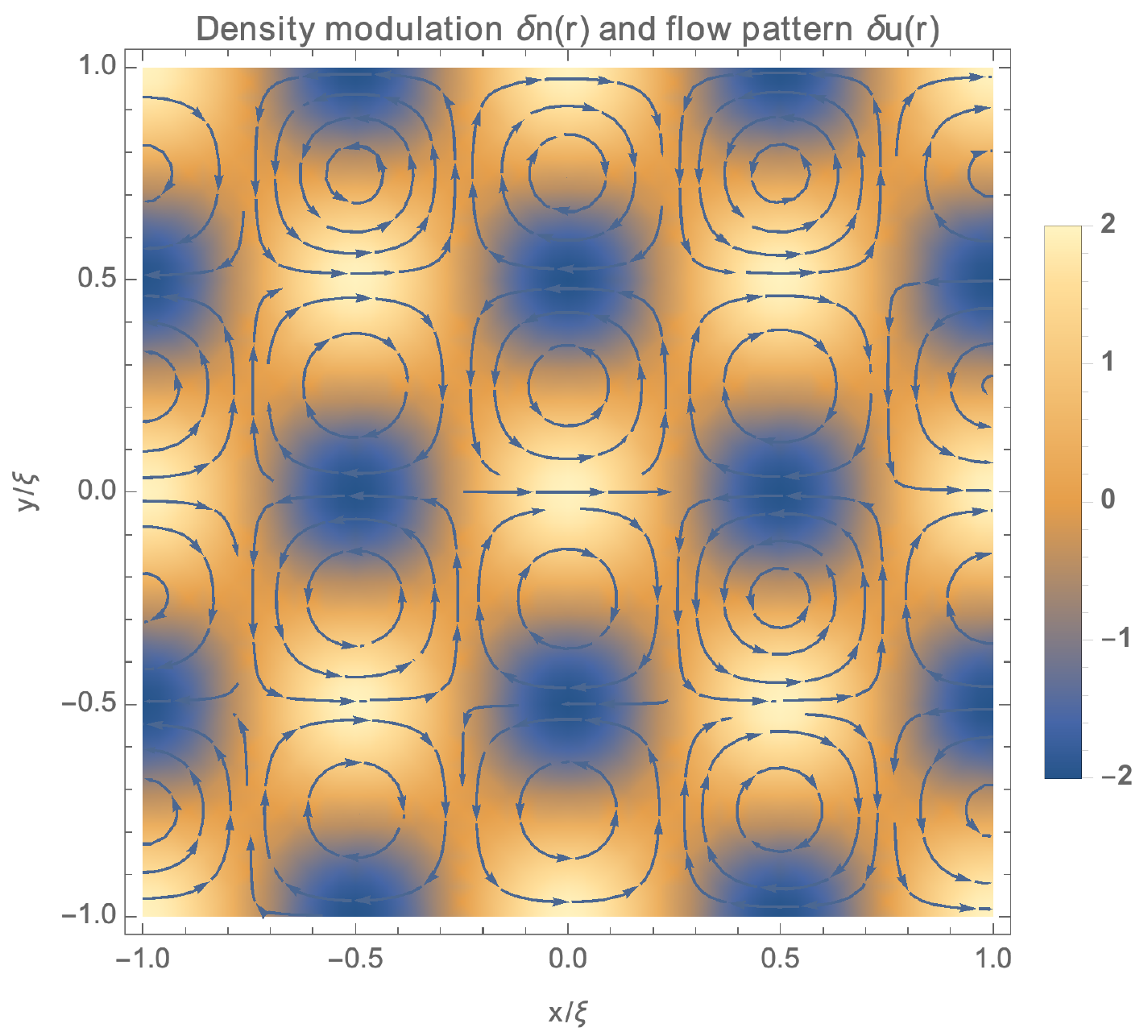}
  \caption{Illustration of the calculated vortical flow pattern induced by the electromotive force in the $x$ direction for a checkerboard density modulation near charge neutrality, $\delta n(\mathbf{r})=\delta n_0\big[\cos(2\pi(x+y)/\xi)+\cos(2\pi(x-y)/\xi)\big]$. The stream-plot flow of velocity field $\delta\bf{u}(\bf{r})$ is indicated by arrows superimposed on the color plot of the density profile $\delta n(\bf{r})$, with positive/negative density shown in yellow/blue, respectively. The correlations of the flow velocity with the density profile produce a macroscopic convective electric current along the $x$ direction, which enhances the conductivity. }
  \label{Fig-model}
\end{figure}

For systems at charge neutrality \cite{Note}, whose resistivity does not vanish even in the pristine disorder-free state \cite{Kashuba,Fritz}, the effect of disorder on the resistivity is not obvious. In previous studies of hydrodynamic electron transport~\cite{Hartnoll,Sachdev,Aleiner,Mirlin}, the system resistivity at charge neutrality was found to be independent of disorder and equal to the intrinsic resistivity of the electron liquid.

Here we show that in the hydrodynamic regime, long-range disorder \emph{increases} the  conductivity of the system at charge neutrality in comparison to the intrinsic conductivity $\sigma_0$ of the electron liquid. The enhancement of the conductivity by disorder is inversely proportional to the shear viscosity of the liquid, and has extremely nonlocal character --- its magnitude grows with  increasing correlation radius of  the disorder potential.  This is in stark contrast to the situation sufficiently far away from charge neutrality~\cite{AKS,Lucas}  where the viscous contributions enhance the resistivity, and their magnitude decreases as the disorder correlation radius grows. This opposite effect on the conductivity is caused by a qualitative difference in the character of the flow caused by the passage of electric current near,  and away from charge neutrality. We show below that near charge neutrality, the flow has a predominantly vortical character whereas away from charge neutrality the flow is predominantly potential.

The vortical character of the induced flow and the reason for the enhancement of conductivity by long-range disorder can be qualitatively understood as follows. Consider a two-dimensional system in the $xy$ plane subject to the electric field $\mathbf{E}_0$. In a pristine system at charge neutrality, electron transport is mediated by the intrinsic conductivity in a stationary liquid. In the presence of disorder, charge neutrality is satisfied only on average, but locally the electron density $\delta n (\mathbf{r})$ is nonzero. The resulting  force density, $\mathbf{F} (\mathbf{r}) =e \mathbf{E}_0 \delta n (\mathbf{r})$, with $e$ being the electron charge, exerted on the fluid  by the electric field must be compensated in a steady state by pressure gradients and viscous stresses arising in the fluid. It is important to note that only the potential part of the external force density, which is caused by density variations along $\mathbf{E}_0$, can be compensated by the pressure gradients. In contrast, the vortical part of the external force that arises from density gradients perpendicular to $\mathbf{E}_0$ must be compensated by viscous stresses. To understand the effect of the density modulations perpendicular to the electric field, let us assume $\mathbf{E}_0$  to be along the $x$ axis and consider for simplicity a density modulation of the form, $\delta n (y) \sim \delta n_0 \cos(y/\xi)$. The force of the electric field $e \mathbf{E}_0 \delta n(y)$ induces an inhomogeneous flow along the $x$ axis, with velocity $u_x = u_0 \cos (y/\xi)$. The velocity amplitude may be estimated by balancing the force of the electric field $eE_0 \delta n$ with the divergence of the viscous stress tensor $\eta \partial_y^2 u_x$, where $\eta$ is the shear viscosity. This yields $u_0 \sim \xi^2 eE_0 \delta n / \eta$. Due to correlations of the induced hydrodynamic velocity with disorder, the hydrodynamic flow gives a nonvanishing contribution to the net current that enhances its magnitude. The corresponding enhancement of the conductivity  may be estimated as
 \begin{equation}\label{eq:delta_sigma}
  \delta \sigma \sim  \frac{e^2 \xi^2 }{\eta}  \left\langle \delta n^2 \right\rangle,
\end{equation}
where $\xi$ is the correlation radius of the disorder potential, and $\langle \ldots \rangle$ denotes averaging over disorder. For a checkerboard pattern of density modulation, which better mimics the isotropic long-range disorder, the calculated vortical flow pattern is illustrated in Fig.~\ref{Fig-model}. The contribution of the vortical velocity variations in the induced hydrodynamic flow to the transport properties of electron systems was ignored in previous  considerations, resulting in disorder-independent conductivity near charge neutrality.

Below we develop a general theory of hydrodynamic transport in a long-range disorder potential without assuming Galilean invariance of the underlying electron liquid. For weak disorder, we obtain general expression for the transport coefficients of the system and apply our results to study thermal and electric transport in graphene devices near charge neutrality. Our consideration shows that the flow near charge neutrality is predominantly vortical, in contrast to nearly potential flow that arises away from charge neutrality. This has a dramatic effect on the transport characteristics of the system. In particular, the estimate in Eq.~\eqref{eq:delta_sigma} for the disorder-induced conductivity enhancement is borne out by the quantitative treatment  presented below.

The paper is organized as follows. In Sec. \ref{sec:hydro}, we present the hydrodynamic description of electron transport in a long-range disorder potential without assuming Galilean invariance of the underlying electron liquid. In Sec. \ref{sec:linear_response}, we apply this description to the study of electron transport in the linear response regime and obtain a general expression for the transport coefficients of the system. In Sec. \ref{sec:long_range}, we simplify these results for the regime of weakly disordered systems with long-range inhomogeneity. Next we apply this general theory to graphene near charge neutrality in Sec.~\ref{sec:graphene}. The summary of our main results is given in Sec. \ref{sec:summary}. To complement our analysis with additional useful details,  various technicalities are delegated to several Appendices.


\section{Hydrodynamic description}
\label{sec:hydro}

The hydrodynamic equations express conservation of the number of particles, energy, and momentum of the electron liquid. Accordingly,  the time derivatives of the number density $n$,  energy density $\epsilon$, and momentum density $p_i$ may be expressed in terms of divergences of the corresponding conserved fluxes, $\mathbf{j}$, $\mathbf{j}_\epsilon$, and $\Pi_{ij}$. In particular, the particle number conservation is
\begin{equation}
\label{eq:continuity_n}
\frac{\partial n}{\partial t}= - \boldsymbol{\nabla}\cdot \mathbf{j}.
\end{equation}
Here and in what follows we denote vector quantities by bold face symbols, and Cartesian indices by Latin subscripts.

In the presence of the external potential $U(\bf{r})$, the evolution equation for momentum density has the form of Newton's second law:
\begin{equation}\label{eq:continuity_p}
\frac{\partial p_i}{\partial t} = - \partial_j \Pi_{ij} -n\, \partial_i (e\phi + U).
\end{equation}
The electric potential $\phi$ here is related to the electron density by the Poisson equation. Its presence in Eq.~\eqref{eq:continuity_p} reflects the flow of momentum of the electron fluid due to long range Coulomb interactions between electrons, whereas the tensor  $\Pi_{ij}$ denotes the local part of the momentum flux.

The final hydrodynamic equation expresses energy conservation of the electron fluid. In addition to the density of particles and momentum of particles, the energy density depends on the entropy density of the liquid. Therefore, in hydrodynamics the energy conservation equation is traditionally replaced  by an equivalent evolution equation for the entropy density~\cite{LL-V6}. The latter may be written in the form,
\begin{equation}
\label{eq:s_dot_source}
 \frac{\partial s}{\partial t} =- \boldsymbol{\nabla}\cdot \mathbf{j}_s +\varsigma,
\end{equation}
where $\mathbf{j}_s$ is the entropy flux,
\begin{equation}
\label{eq:j_S_def}
\mathbf{j}_s= \frac{1}{T} \left[ \mathbf{j}_\epsilon  - (\mu + e\phi + U)\, \mathbf{j} \right],
\end{equation}
and  $\varsigma$ denotes the local rate of entropy production due to electron-electron collisions.

A crucial ingredient of the hydrodynamic approach is the assumption of local thermal equilibrium of the electron liquid. Accordingly,  the state of the liquid is characterized by the local equilibrium parameters: temperature $T$, chemical potential $\mu$,  and the hydrodynamic velocity $\mathbf{u}$, whose values are determined by the local densities of conserved quantities. Furthermore, the fluxes of conserved quantities are described in terms of the gradient expansion in equilibrium parameters.

Being interested in the linear response properties of the system we may write this gradient expansion in the form,
\begin{subequations}\label{eq:constitutive}
  \begin{eqnarray}
    &&\mathbf{j} = n \mathbf{u} + \mathbf{j}',  \label{eq:constitutive_j}\\
    &&\mathbf{j}_\epsilon = ( w + ne\phi + n U) \mathbf{u} + \mathbf{j}_\epsilon', \label{eq:constitutive_j_epsilon} \\
    &&\Pi_{ij} = P \delta_{ij}  -\sigma'_{ij}, \label{eq:constitutive_Pi}
  \end{eqnarray}
\end{subequations}
where $n$ is the electron number density, $w$ is the enthalpy density,  and $P$ denotes the local  pressure. The first terms on the right-hand side in the above equations denote the equilibrium components of fluxes of conserved quantities. The primed quantities denote the dissipative fluxes that are proportional to gradients of the equilibrium parameters. In particular, $\sigma'_{ij}$ denotes the viscous stress tensor,
\begin{equation}\label{eq:viscous_stress}
\sigma_{ij}' = \eta (\partial_i u_j + \partial_j u_i) + \left(\zeta- \frac{2}{d}\eta\right) \delta_{ij}
\partial_k u_k,
\end{equation}
where $\eta$ and $\zeta$ are, respectively, shear and bulk viscosities, and $d$ is the dimensionality of space.

The entropy production rate $\varsigma$ in Eq.~\eqref{eq:s_dot_source} can be expressed in terms of the dissipative fluxes of conserved quantities in the form (see Appendix \ref{sec:entropy_production} for a derivation)
\begin{equation}\label{eq:s_dot_local_currents}
 \varsigma=  - \frac{ \mathbf{j}'_s \cdot \boldsymbol{\nabla} T  +    \mathbf{j}' \cdot\boldsymbol{\nabla} \left( \mu + e\phi + U\right)  + u_i \partial_j  \sigma'_{ij}   }{T}  .
\end{equation}
Here $\mathbf{j}'_s$ denotes the dissipative part of the entropy flux in Eq.~\eqref{eq:j_S_def}, which is defined as
\begin{equation}
\label{eq:j_S'_def}
\mathbf{j}'_s \equiv \mathbf{j}_s - s \mathbf{u} = \frac{\mathbf{j}_\epsilon' - (\mu + e \phi + U) \mathbf{j}'}{T}.
\end{equation}
The last equality follows from Eqs.~\eqref{eq:constitutive_j} and \eqref{eq:constitutive_j_epsilon} and the thermodynamic relation $w=n\mu + Ts$.

The hydrodynamic equations need to be supplemented by the  constitutive relations for the relevant fluxes  in terms of the gradients of the equilibrium parameters. To keep subsequent expressions more compact, it is convenient to combine the particle and entropy fluxes into a two-component column vector:
\begin{equation}\label{eq:J_column}
  \vec{\mathbf{J}}= \left(
                      \begin{array}{l}
                        \mathbf{j} \\
                        \mathbf{j}_s \\
                      \end{array}
                    \right).
\end{equation}
Throughout the paper, we indicate two-component column vector quantities by arrows above them, and use bold face letters to denote the usual spatial vectors. Following the conventions of Ref.~\cite{LL-V5}, we denote densities of thermodynamic variables by $x_i$ and the corresponding thermodynamically conjugate quantities by $X_i$.
Introducing the column vector notations
\begin{equation}\label{eq:x_X_def}
 \vec{x}= \left(
    \begin{array}{c}
      n \\
      s \\
    \end{array}
  \right), \quad \vec{\mathbf{X}}= \left(
    \begin{array}{c}
      - e\boldsymbol{\mathcal{E}} \\
       \boldsymbol{\nabla} T \\
    \end{array}
  \right),
\end{equation}
where
\begin{equation}\label{eq:EMF_def}
  e  \boldsymbol{\mathcal{E}}\equiv -\boldsymbol{\nabla} \left( \mu + e\phi + U\right)
\end{equation}
is  the electromotive force,  we can write the constitutive relations for the particle and entropy currents in the form
\begin{eqnarray}\label{eq:column_constitutive}
 \vec{\mathbf{J}}= \vec{x}\mathbf{u} - \hat{\Upsilon}\vec{\mathbf{X}}.
\end{eqnarray}
Here $\hat{\Upsilon}$ is the matrix of kinetic coefficients that characterizes the dissipative properties of the electron liquid. It is given by
\begin{equation}
 \label{eq:gamma_def}
 \quad \hat{\Upsilon}= \left(
                                 \begin{array}{cc}
                                   \sigma/e^2 & \gamma/T \\
                                   \gamma/T & \kappa/T  \\
                                 \end{array}
                               \right),
\end{equation}
where $\kappa$ is the thermal conductivity, $\sigma$ is the intrinsic conductivity, and $\gamma$ is the thermoelectric coefficient of the electron liquid. Throughout the paper, we set the Boltzmann and Planck constants to unity, $k_B=\hbar=1$.

The system of hydrodynamic equations and constitutive relations presented in this section does not assume Galilean invariance and provides a general description of the flow of electron liquid in an external potential at small velocities. For  Galilean-invariant liquids $\mathbf{j}=n \mathbf{u}$, and  $\sigma=\gamma=0$. In this case, the second term in the numerator of the right-hand side of Eq.~\eqref{eq:s_dot_local_currents} vanishes, and  Eq.~\eqref{eq:s_dot_local_currents} reproduces the well-known result for the energy dissipation rate in Galilean-invariant liquids \cite{LL-V6}.


\section{Electron transport in linear response}\label{sec:linear_response}

We now use the hydrodynamic description presented in Sec.~\ref{sec:hydro} to study  electron transport in the linear response regime. We consider  the stationary (\emph{dc}) situation, and, being motivated by applications to graphene systems, focus on the two-dimensional (2D) geometry.

In linear response, we may neglect entropy production and set $\varsigma \to 0$ in Eq.~\eqref{eq:s_dot_source}. Then in the stationary case, the continuity equation \eqref{eq:continuity_n} and the entropy evolution equation \eqref{eq:s_dot_source} reduce to the continuity equations for the
column vector current:
\begin{equation}\label{eq:column_current_continuity}
 \boldsymbol{\nabla} \cdot\left( \vec{x}\mathbf{u} - \hat{\Upsilon}\vec{\mathbf{X}} \right)=0.
\end{equation}
Using Eq.~\eqref{eq:constitutive_Pi} along with an explicit form of the stress tensor from Eq.~\eqref{eq:viscous_stress}, and the thermodynamic relation
$\boldsymbol{\nabla} P = n \boldsymbol{\nabla} \mu + s \boldsymbol{\nabla} T$, the momentum balance equation \eqref{eq:continuity_p} for a 2D case may be expressed as
\begin{equation}\label{eq:momentum_balance}
  e n \boldsymbol{\mathcal{E}}  - s \bm{\nabla} T + \left(\bm{\nabla} \cdot  \eta \bm{\nabla} \right)\mathbf{u} + \bm{\nabla} \left[\zeta \left(\boldsymbol{\nabla} \cdot\mathbf{u}\right)\right]= 0,
\end{equation}
where the electromotive force $e \boldsymbol{\mathcal{E}} $ is given by Eq.~(\ref{eq:EMF_def}).
In the column vector notations, the force balance relation \eqref{eq:momentum_balance} is cast in the form
\begin{equation}\label{eq:force_balance_column}
  -\vec{x}^{\mathbb{T}}\vec{\bf X} + \left(\bm{\nabla} \cdot \eta \bm{\nabla} \right)\mathbf{u} + \bm{\nabla} \left[\zeta \left(\boldsymbol{\nabla} \cdot\mathbf{u}\right)\right]= 0,
\end{equation}
where we used the standard rules for multiplying the column vector $\vec{x}^{\mathbb{T}}  = (n, s) $, with the superscript $\mathbb{T}$ indicating the transposition, and the row vector $\vec{\bf{X}}$.

In thermal equilibrium, the hydrodynamic velocity vanishes. Since both the temperature and the electrochemical potential $\mu + e\phi + U$ are spatially uniform, the thermodynamic force $\vec{\bf X}$ vanishes. As a result, Eqs.~\eqref{eq:column_current_continuity} and \eqref{eq:force_balance_column} are trivially satisfied.

Away from equilibrium, one needs to find a nonvanishing spatial distribution of the hydrodynamic velocity $\mathbf{u} (\bf{r})$ and the thermodynamic force $\vec{\bf X}(\bf{r})$ that solves the system of equations \eqref{eq:column_current_continuity} and  \eqref{eq:force_balance_column}. In linear response, the number density and entropy density in the column vector $\vec{x}(\bf{r})$, as well as the dissipative coefficients of the liquid, $\hat{\Upsilon}$, $\eta$ and $\zeta$ are given by their equilibrium values. As follows from the definition in Eq.~\eqref{eq:x_X_def}, the column vector field $\vec{\bf X}$ must be purely potential. Thus the problem of determination of transport properties of the system reduces to solving the linear flow problem and calculating the net particle and entropy fluxes averaged through the system. For this purpose, we combine the macroscopic densities of particle and entropy flux into a column vector current,
\begin{equation}
 \label{eq:averaged_current_spatial}
  \langle \vec{\bf J}\rangle=\left\langle \vec{x} (\r) \mathbf{u} (\r) - \hat{\Upsilon} (\r)  \vec{\mathbf{X}} (\r) \right\rangle,
\end{equation}
where $\langle \ldots \rangle \equiv \frac{1}{A} \int d \bf{r}\, (\ldots) $ denotes spatial average over the 2D system with area $A$.

Mathematically, this problem is similar to the problem of finding the spatial distribution of current  in an inhomogeneous conductor, where one needs to determine two vector fields: one divergence-free, current density $\mathbf{j}(\mathbf{r})$, and one purely potential, electric field $\mathbf{E}(\mathbf{r})$, which are related by the position-dependent conductivity $\mathbf{j}(\bf{r})= \sigma (\bf{r}) \mathbf{E}(\bf{r})$ \cite{Dykhne}.

In the framework of the present approach, the disorder potential manifests itself via the spatial dependence of the equilibrium number and entropy densities, $n(\bf{r})$ and $s(\bf{r})$,  the matrix of kinetic coefficients of the liquid,  $\hat{\Upsilon}(\bf{r})$, and the viscosities $\eta(\bf{r})$ and $\zeta(\bf{r})$. Below we assume that these quantities are weakly inhomogeneous and use perturbation theory in disorder to derive analytical results for the macroscopic thermoelectric conductivity matrix $\hat{\Upsilon}_{\rm{e}}$. The latter relates the macroscopic particle and entropy  currents, $\langle \vec{\mathbf{J}} \rangle = \langle (\mathbf{j},\mathbf{j}_s)^{\mathbb{T}}\rangle$, to the average electric field and the temperature gradient in the system, $\vec{\mathbf{X}}_0\equiv \langle \vec{\mathbf{X}} \rangle$:
\begin{equation}\label{eq:Upsilon_eff_def}
   \langle \vec{\bf J}\rangle  = \hat{\Upsilon}_{\rm{e}} \vec{\mathbf{X}}_0.
\end{equation}
Its inverse,  $\hat{\varrho}=[\hat{\Upsilon}_{\rm{e}} ]^{-1}$, defines the macroscopic thermoelectric resistivity matrix.

The presentation in this section is organized as follows.  We begin by considering a uniform liquid in Sec.~\ref{sec:uniform}. Then in Sec.~\ref{sec:perturbation}, we develop perturbation theory about the uniform solution. Finally, in Sec.~\ref{sec:matrix} we obtain general perturbative expressions for the transport coefficients of the system.


\subsection{Uniform liquid}
\label{sec:uniform}

In a uniform liquid with number density $n_0$ and entropy density $s_0$ the charge and energy transport can proceed in two ways: (i) via a hydrodynamic flow with a uniform hydrodynamic velocity $\mathbf{u}_0$, and (ii) by transport relative to the liquid driven by the thermodynamic forces $\vec{\bf X}_0$. The momentum balance condition Eq. (\ref{eq:momentum_balance}) implies that for a uniform liquid a steady  state can exist only if the uniform electric field $\boldsymbol{\mathcal{E}}_0$ and temperature gradient $\boldsymbol{\nabla} T_0$ satisfy the relation
\begin{equation}
e n_0 \boldsymbol{\mathcal{E}}_0  = s_0 \boldsymbol{\nabla}T_0.
\end{equation}
Under this condition, the force due to the external electric field is balanced by the pressure gradient, so no acceleration of the liquid occurs.
The corresponding column force $\vec{\bf X}_0$ may be expressed in the form
\begin{equation}\label{eq:X_0_def}
  \vec{\mathbf{X}}_0
  = \frac{e \boldsymbol{\mathcal{E}}_0}{ s_0}
  \left(
    \begin{array}{c}
      -  s_0 \\
      n_0   \\
    \end{array}
  \right)
  =-\frac{i \hat{\tau}_y \vec{x}_0}{s_0} \, e\boldsymbol{\mathcal{E}}_0,
\end{equation}
where  the column vector $\vec{x}_0$ describes the particle and entropy density in the uniform state, see Eq.~\eqref{eq:x_X_def}, and $\hat{\tau}_y$ is the Pauli matrix acting in the $2\times 2$ column-vector space. Thus, the steady-state current may be written in the  form
\begin{equation}\label{eq:solution_uniform}
  \vec{\mathbf{J}}_0=\vec{x}_0\mathbf{u}_0 +\hat{\Upsilon}_0\, \frac{i \hat{\tau}_y \vec{x}_0}{s_0}  e\calE_0.
\end{equation}
The first term describes the dissipationless transport in the hydrodynamic mode caused by the uniform flow of the liquid, and the second term describes charge and energy transport that occurs relative to the liquid at rest. We refer to the latter as the relative transport mode below.  For example, at charge neutrality, charge transport is entirely due to the relative mode, while the hydrodynamic mode corresponds to convective heat flow and does not contribute to electric current. The entropy production obtained from Eq. \eqref{eq:s_dot_local_currents} by integrating over space,
\begin{equation}\label{eq:entropy production_uniform}
  T \dot{S}= \int d \r
 \left( \frac{e \calE_0}{s_0}\right)^2
 (i\hat{\tau}_y \vec{x}_0)^{\mathbb{T}}\hat{\Upsilon}_0(i\hat{\tau}_y\vec{x}_0),
\end{equation}
is entirely due to the transport in the relative mode.

The dissipationless transport mode exists only in the absence of disorder. At finite disorder, the relative and the hydrodynamic modes mix. This mixing and the viscous stress arising in the inhomogeneous flow cause additional dissipation. At weak disorder, this dissipation is expected to be especially significant for the hydrodynamic transport mode described by the first term in Eq.~\eqref{eq:solution_uniform}. However, as we will show below, transport in the relative (on average) mode  can also be significantly modified by the mixing between the relative and hydrodynamic modes. In particular, at charge neutrality the conductivity of the system may significantly deviate from the intrinsic conductivity of the electron liquid. As we show in Sec.~\ref{sec:graphene}, this can occur even at weak disorder provided its correlation radius is sufficiently long.


\subsection{Perturbation theory in disorder}
\label{sec:perturbation}

We now consider the situation in which the parameters $\vec{x}(\r)$ and $\hat{\Upsilon}(\r)$ describing the equilibrium state of the liquid are weakly inhomogeneous functions of position $\r$  (as we will show, to second order in inhomogeneity the spatial variations of the viscosities of the electron liquid may be neglected). To this end, we write
\begin{equation}
\vec{x}(\r) = \vec{x}_0 +\delta \vec{x}(\r), \quad  \hat{\Upsilon}(\r) = \hat{\Upsilon}_0 + \delta \hat{\Upsilon}(\r),
\end{equation}
where $\vec{x}_0$ and $\hat{\Upsilon}_0$ denote the uniform components of and $\vec{x}(\r)$ and $\hat{\Upsilon}(\r)$, respectively, while  $\delta \vec{x}(\r) \ll \vec{x}_0$ and $\delta \hat{\Upsilon}(\r) \ll \hat{\Upsilon}_0$ denote their spatial variations.
We assume that the resulting hydrodynamic velocity $\mathbf{u}(\r)$ and the thermodynamic force $\vec{\bf X} (\r)$ are also nearly homogeneous,
\begin{equation}
\mathbf{u}(\r) = \mathbf{u}_0 +\delta \mathbf{u}(\r), \quad  \vec{\bf X} (\r) = \vec{\bf X}_0 + \delta \vec{\bf X} (\r),
\end{equation}
with $\delta \mathbf{u}(\r) \ll  \mathbf{u}_0$ and $\delta \vec{\bf X} (\r) \ll \vec{\bf X}_0$.

The inhomogeneous components of the hydrodynamic velocity $\delta\mathbf{u}(\r)$ and the driving force $\delta \vec{\bf X} (\r)$ may be determined using   perturbation theory about the uniform solution. The hydrodynamic equations to be solved consist of the continuity equation for the column current, Eq.~\eqref{eq:column_current_continuity}, and the momentum balance equation, Eq.~\eqref{eq:force_balance_column}.

The solution strategy can be summarized as follows. We determine the  inhomogeneous part of the flow velocity field $\delta \mathbf{u}$ and forces $\delta \vec{\mathbf{X}}$ in terms of their uniform counterparts to linear order in $\delta \vec{x}$ and $\delta \hat{\Upsilon}$. This enables us to express the spatial average of the currents Eq. \eqref{eq:averaged_current_spatial} in terms of $\vec{\bf{X}}_0$ and $\bf{u}_0$ to second order accuracy in inhomogeneity. Furthermore, the spatial average of the momentum balance equation \eqref{eq:force_balance_column} imposes a linear relation between $\vec{\bf{X}}_0$ and $\bf{u}_0$. Using this relation we express the macroscopic currents in the form of Eq. \eqref{eq:Upsilon_eff_def} and thus determine $ \hat{\Upsilon}_{\rm{e}}$.

To implement this program, we switch to the Fourier representation, defining the Fourier amplitudes of various quantities $O (\mathbf{r})$ in the standard way:
\begin{equation}\label{eq:Fourier_def}
  O_{\q} = \int d\mathbf{r}\, O (\mathbf{r})  e^{- i \q \cdot \mathbf{r}}.
\end{equation}
To linear order accuracy in the perturbations, we get for the inhomogeneous Fourier components of the column current
\begin{equation}\label{eq:perturbation}
  \vec{\mathbf{J}}_\q=\vec{x}_\q\mathbf{u}_0 +
   \vec{x}_0\mathbf{u}_\q  - \hat{\Upsilon}_\q  \vec{\mathbf{X}}_0
   -\hat{\Upsilon}_0\vec{\mathbf{X}}_\q.
\end{equation}
The continuity equations for the heat and particle currents, Eq.~\eqref{eq:column_current_continuity}, then become
\begin{equation}\label{eq:continuity_perturbation}
   \vec{x}_0 (\q \cdot \mathbf{u}_\q)-\hat{\Upsilon}_0\, (\q \cdot \vec{\bf X}_\q)=-\vec{x}_\q (\q \cdot \mathbf{u}_0) +
   \hat{\Upsilon}_\q\, (\q \cdot \vec{\bf X}_0).
\end{equation}
The force balance equation \eqref{eq:force_balance_column} imposes the following relation on the  Fourier components of velocity and thermodynamic forces with $\q \neq 0$:
\begin{equation}
\label{eq:u_q_X_column}
\eta q^2 \mathbf{u}_\q
+ \zeta  \q (\q \cdot \mathbf{u}_\q)  + \vec{x}_0^{\mathbb{T}} \vec{\bf X}_\q +\vec{x}_\q^{\mathbb{T}} \vec{\bf X}_0 =0 .
\end{equation}

The system of equations \eqref{eq:continuity_perturbation} and \eqref{eq:u_q_X_column} determines the inhomogeneous components of the hydrodynamic velocity, $\mathbf{u}_\q$,  and electric field/temperature gradient, $\vec{\bf X}_\q$, in terms of the macroscopic hydrodynamic velocity, $\mathbf{u}_0$, and macroscopic electric field/temperature gradient, $\vec{\bf X}_0$. However, since the macroscopic flow is characterized by only two macroscopic currents (particle and entropy flux, $\langle \mathbf{j} \rangle, \langle \mathbf{j}_s \rangle$), the average velocity $\mathbf{u}_0$ is not independent from $\vec{\bf X}_0$.  The relation between them can be obtained by considering the uniform Fourier component of the  force balance equation \eqref{eq:force_balance_column}, which can be written in the form
\begin{equation}\label{eq:force_balance_column_uniform}
  -\vec{x}^{\mathbb{T}}_0\vec{\bf X}_0 - \left\langle \delta \vec{x}^{\mathbb{T}} \delta \vec{\bf{X}}   \right\rangle  = 0.
\end{equation}
Substituting the solutions of Eqs.~\eqref{eq:continuity_perturbation} and \eqref{eq:u_q_X_column} into Eq. \eqref{eq:force_balance_column_uniform}, one can express average velocity $\mathbf{u}_0$ in terms of $\vec{\bf X}_0$ to second order accuracy in disorder.

The solution of the system of linear equations \eqref{eq:continuity_perturbation}--\eqref{eq:u_q_X_column} is given by
\begin{subequations}
\label{eq:solution_summary}
\begin{align}
\label{eq:X_q_result}
&\vec{\mathbf{X}}_\q = \frac{\q}{q^2 \lambda_q}   \hat{\Upsilon}_0^{-1}
\left\{\left[ \lambda_q - \vec{x}_0 \otimes \vec{x}_0^{\mathbb{T}} \hat{\Upsilon}_0^{-1}\right] \vec{x}_\q  (\q \cdot \mathbf{u}_0) \right. \nonumber \\
  & +\left. \left[ \left(  \vec{x}_0 \otimes \vec{x}_0^{\mathbb{T}} \hat{\Upsilon}_0^{-1} - \lambda_q \right)\hat{\Upsilon}_\q
   - \vec{x}_0  \otimes \vec{x}_\q^{\mathbb{T}} \right]
   (\q \cdot \vec{\mathbf{X}}_0) \right\},\\
 \label{eq:u_transverse_summary_sol}
&\mathbf{u}_\q^t =
  - \frac{1}{ \eta q^2}  \, \vec{x}_\q^{\mathbb{T}}\,    \left[  \vec{\bf{X}}_0 -  \frac{\q\,( \q \cdot  \vec{\mathbf{X}}_0) }{q^2} \right] , \\
  \label{eq:u_longitudinal_summary_sol}
&\mathbf{u}_\q^l=-\frac{  \vec{x}_0^{\mathbb{T}} \,
 \hat{\Upsilon}_0^{-1}      \vec{x}_\q
}{\lambda_q} \, \frac{\q \, (\q \cdot \mathbf{u}_0)}{q^2}\nonumber \\
&+  \frac{\left(
\vec{x}_0^{\mathbb{T}}  \hat{\Upsilon}_0^{-1}  \hat{\Upsilon}_\q
- \vec{x}_\q^{\mathbb{T}}
\right)}{\lambda_q}
\, \frac{\q ( \q \cdot \vec{\mathbf{X}}_0) }{q^2}.
\end{align}
\end{subequations}
Here $\mathbf{u}_\q^t$ and $\mathbf{u}_\q^l$ denote, respectively, the transverse and longitudinal components of the hydrodynamic velocity and $\lambda_q$ is a function of momentum $q$ given by
\begin{equation}
\label{eq:lambda_q_def} \lambda_q = (\eta + \zeta) q^2 +\vec{x}_0^{\mathbb{T}} \,
 \hat{\Upsilon}_0^{-1}      \vec{x}_0.
\end{equation}
We indicate transposition of column vectors by superscript $\mathbb{T}$ and use the standard notation for the direct product of two vectors $\vec{a}\otimes\vec{b}^{\mathbb{T}}$ that defines a corresponding matrix.

Substituting the result Eq. \eqref{eq:X_q_result} for $\vec{\bf{X}}_{\bf{q}}$ into the macroscopic momentum balance equation \eqref{eq:force_balance_column_uniform}, we obtain the following relation between $\mathbf{u}_0$ and $\vec{\bf X}_0$:
\begin{align}
\label{eq:u_0_X_0_relation}
  &\vec{x}^{\mathbb{T}}_0\vec{\bf X}_0 = -\frac{ \mathbf{u}_0}{2 } \int_{\bf{q}}  \left[
  \vec{x}_{-\q}^{\mathbb{T}}  \hat{\Upsilon}_0^{-1} \vec{x}_\q  - \frac{ |\vec{x}_0^{\mathbb{T}} \hat{\Upsilon}_0^{-1} \vec{x}_\q|^2 }{ \lambda_q} \right] \nonumber \\
  &  + \frac{1}{2} \int_{\bf{q}} \left[  \vec{x}_{-\q}^{\mathbb{T}} \hat{\Upsilon}_0^{-1} \hat{\Upsilon}_\q
  \right]\, \vec{\bf X}_0
  \nonumber \\
  & - \frac{1}{2} \int_{\bf{q}}  \left[ \vec{x}_{-\q}^{\mathbb{T}}   \hat{\Upsilon}_0^{-1}      \vec{x}_0\right]
   \left[ \vec{x}_0^{\mathbb{T}} \hat{\Upsilon}_0^{-1} \hat{\Upsilon}_\q -
   \vec{x}_\q^{\mathbb{T}}\right]
     \frac{\vec{\bf X}_0}{\lambda_q}
\end{align}
where
the factor $1/2$ arises from the projection into the longitudinal and transverse components that is specific to two dimensions, and
consequently, $\int_\q \ldots  = \int \frac{d^d q}{(2\pi)^d} \ldots $, denotes an integral over the wave vectors in $d=2$.

Equation \eqref{eq:u_0_X_0_relation} expresses force balance on spatial scales large in comparison to the disorder correlation radius. Comparing  this equation with Eq.~\eqref{eq:force_balance_column}, we see  that the last two terms on the right-hand side of Eq.~\eqref{eq:u_0_X_0_relation} describe renormalization of the effective densities of particle number and entropy by disorder. In contrast, the first term on the right-hand side of Eq. \eqref{eq:u_0_X_0_relation} describes the emergent friction force due to the macroscopic flow of the electron liquid, $\mathbf{F}_{f}  = - k \mathbf{u}_0$. Thus the spatially uniform part of the force balance equation, Eq.~\eqref{eq:u_0_X_0_relation}, can be written in the form
\begin{equation}\label{eq:momentum_balance_renormalized}
  - \vec{x}^{\mathbb{T}}_{\rm{e}} \vec{\bf X}_0 - k \mathbf{u}_0 = 0.
\end{equation}
Here the column-vector $\vec{x}_{\rm{e}}$ describing disorder-renormalized particle and entropy densities is given by
\begin{align}
\label{eq:x_eff}
  &\vec{x}_{\rm{e}} = \vec{x}_0 -
    \frac{1}{2}\int_{\bf{q}} \left[ \hat{\Upsilon}_{-\q}  \hat{\Upsilon}_0^{-1}   \vec{x}_{\q}\right.\nonumber \\
  &-\left.\lambda^{-1}_q\left(  \hat{\Upsilon}_{-\q} \hat{\Upsilon}_0^{-1} \vec{x}_0  -
   \vec{x}_{-\q}\right) \, \vec{x}_0^{\mathbb{T}} \,  \hat{\Upsilon}_0^{-1} \vec{x}_{\q}
    \right],
\end{align}
and the friction coefficient $k$ is given by
\begin{equation}\label{eq:friction_coefficient}
  k     = \frac{1}{2}\int_{\bf{q}}  \vec{x}_{-\q}^{\mathbb{T}}  \hat{K}_ q \vec{x}_\q ,
\end{equation}
where we introduced the matrix
\begin{equation}\label{eq:K_hat}
 \hat{K}_{q}  = \hat{\Upsilon}_0^{-1} - \frac{1}{\lambda_q} \hat{\Upsilon}_0^{-1}   \vec{x}_0\otimes  \vec{x}_0^{\mathbb{T}} \hat{\Upsilon}_0^{-1} .
\end{equation}
The friction coefficient $k$ is positive definite, as we show below. Using the notations from Eqs. \eqref{eq:x_eff} and \eqref{eq:friction_coefficient}, the spatial average of the hydrodynamic velocity $\mathbf{u}_0$ can be expressed in terms of the average electric field and temperature gradient $ \vec{\mathbf{X}}_0$ from Eq. \eqref{eq:momentum_balance_renormalized}.


\subsection{Macroscopic thermoelectric conductivity matrix}\label{sec:matrix}

We are now in a position to evaluate the transport coefficients of the system. This can be done in two equivalent ways: (i)  by expressing the macroscopic particle and entropy current $\langle \vec{\bf J}\rangle$ in terms $\vec{\bf{X}}_0$, or (ii) by evaluating the entropy production rate in terms of  $\vec{\bf{X}}_0$. We continue by following the first route. The second approach leads to the same result, and is described in Appendix \ref{sec:resistivity_matrix}.

To this end, we express the macroscopic particle and entropy currents, $\langle \vec{\bf J}\rangle$, in terms of $\vec{\bf{X}}_0$ and $\mathbf{u}_0$.  Rewriting next Eq.~\eqref{eq:averaged_current_spatial} in terms of the integrals in reciprocal space of wave numbers,
\begin{equation}
\langle \vec{\bf J}\rangle =  \vec{x}_0 \mathbf{u}_0 - \hat{\Upsilon}_0 \vec{\bf{X}}_0  + \int_{\bf{q}} \left(\vec{x}_{-\q} \mathbf{u}_\q  - \hat{\Upsilon}_{-\q} \vec{\mathbf{X}}_\q\right),
\end{equation}
and using the linear response solutions from Eq.~\eqref{eq:solution_summary}, we can express the macroscopic particle and entropy currents in the systems in terms of $\vec{\bf{X}}_0$ and $\mathbf{u}_0$ in the form
\begin{eqnarray}\label{eq:average_current_u_X}
 \langle \vec{\bf J}\rangle  & = & \vec{x}_{\rm{e}} \mathbf{u}_0  -  \hat{\Upsilon}_0 \vec{\mathbf{X}}_0 \nonumber  \\
 && -   \frac{1}{2} \int_{\bf{q}} \left[\frac{1}{\lambda_q} \hat{\Upsilon}_{-\q}
  \hat{\Upsilon}_0^{-1}
 \left(  \vec{x}_0 \otimes \vec{x}_0^{\mathbb{T}} \hat{\Upsilon}_0^{-1} - \lambda_q \right)\hat{\Upsilon}_\q  \right.
 \nonumber \\
 &&- \left.
  \frac{1}{\lambda_q} \left(  \vec{x}_{-\q} \otimes
\vec{x}_0^{\mathbb{T}}  \hat{\Upsilon}_0^{-1}  \hat{\Upsilon}_\q  +  \hat{\Upsilon}_{-\q}
  \hat{\Upsilon}_0^{-1}  \vec{x}_0  \otimes \vec{x}_\q^{\mathbb{T}} \right)
   \right. \nonumber \\
 && \left.
 +\left( \frac{ 1 }{\eta q^2 } +\frac{1}{\lambda_q} \right)  \vec{x}_{-\q} \otimes \vec{x}_\q^{\mathbb{T}}
   \right]   \vec{\mathbf{X}}_0.
\end{eqnarray}
Here the contributions proportional to $\mathbf{u}_0$ are expressed in terms of the column vector of disorder-renormalized densities defined in Eq.~\eqref{eq:x_eff}. Finally, substituting $\mathbf{u}_0$ from Eq. \eqref{eq:momentum_balance_renormalized} into the last equation, we obtain Eq.~\eqref{eq:Upsilon_eff_def} with the matrix of effective kinetic coefficients of the medium $\hat{\Upsilon}_{\rm{e}}$  given by
\begin{eqnarray}\label{eq:Upsilon_eff_general}
 \hat{\Upsilon}_{\rm{e}} & = &\frac{1}{k}\big(\vec{x}_{\rm{e}} \otimes \vec{x}_{\rm{e}}^{\mathbb{T}}\big)  +   \hat{\Upsilon}_0 \nonumber  \\
 &&  +  \frac{1}{2}  \int_{\bf{q}}\left[  \frac{1}{\lambda_q} \hat{\Upsilon}_{-\q}
  \hat{\Upsilon}_0^{-1}
 \left(  \vec{x}_0 \otimes \vec{x}_0^{\mathbb{T}}\hat{\Upsilon}_0^{-1} - \lambda_q \right)\hat{\Upsilon}_\q  \right.
 \nonumber \\
 &&-   \left.
  \frac{1}{\lambda_q} \left(  \vec{x}_{-\q} \otimes
\vec{x}_0^{\mathbb{T}}  \hat{\Upsilon}_0^{-1}  \hat{\Upsilon}_\q  +  \hat{\Upsilon}_{-\q}
  \hat{\Upsilon}_0^{-1}  \vec{x}_0  \otimes \vec{x}_\q^{\mathbb{T}} \right)
  \right. \nonumber \\
 &&  + \left.
 \left( \frac{ 1 }{\eta q^2 } +\frac{1}{\lambda_q} \right)  \vec{x}_{-\q} \otimes \vec{x}_\q^{\mathbb{T}}
   \right]  .
\end{eqnarray}
The matrix in the above expression is symmetric in agreement with the Onsager symmetry principle of kinetic coefficients.

Let us now summarize the results of this subsection. Equation \eqref{eq:Upsilon_eff_general}, together with Eqs.~\eqref{eq:lambda_q_def} and \eqref{eq:x_eff}--\eqref{eq:K_hat}, express the thermoelectric conductivity matrix of the system in terms of the position-dependent matrix of kinetic coefficients of the liquid, $\hat{\Upsilon} (\mathbf{r}) $, and densities of particles and entropy, $\vec{x}(\mathbf{r}) $, in the equilibrium state. Note that the conductivity of the system cannot be expressed in terms of local fluctuations of $ \vec{x}(\mathbf{r}) $ and $\hat{\Upsilon}(\mathbf{r}) $.


\section{Long range disorder}
\label{sec:long_range}

In this section we tailor the above general framework to an experimentally relevant case of long-range disorder. This situation can be realized in high mobility semiconductor quantum wells with modulation doping and boron nitride encapsulated graphene devices
\cite{Gao,Bandurin-1,Crossno,Ghahari,Kumar,Morpurgo,Bandurin-2,Berdyugin}.

In this case, the expressions for the transport coefficients of the system obtained in the previous section may be simplified significantly by selecting the terms that scale as leading powers of the correlation radius $\xi$ of the disorder potential.  More specifically, we assume that the correlation radius satisfies the condition
\begin{equation}\label{eq:locality_condition}
  \varepsilon \equiv   \frac{1}{\xi^2}\frac{  \eta + \zeta }{\vec{x}_0^{\mathbb{T}} \,
 \hat{\Upsilon}_0^{-1}      \vec{x}_0} \ll 1
\end{equation}
and obtain the transport coefficients of the system to leading order in $\varepsilon$.

It is important to note that when the condition Eq. \eqref{eq:locality_condition} is satisfied, the transverse component of the velocity in Eq.~\eqref{eq:u_transverse_summary_sol}, which corresponds to the vortical flow, exceeds the last term in Eq.~\eqref{eq:u_longitudinal_summary_sol}. The latter corresponds to the potential component of the flow caused by the thermodynamic forces $\vec{\mathbf{X}}_0$. This implies that in transport measurements dominated by the relative mode, for which the  macroscopic flow velocity $\mathbf{u}_0$ is small, the inhomogeneous part of the flow induced on a spatial scale of order of the correlation radius $\xi$  is primarily vortical. In particular, this situation is realized in charge transport near the neutrality point, as was qualitatively discussed in the Introduction. For transport measurements in which the macroscopic hydrodynamic flow characterized by the velocity $\mathbf{u}_0$ plays a substantial role, the first term in Eq.~\eqref{eq:u_longitudinal_summary_sol} may exceed the vortical contribution Eq. \eqref{eq:u_transverse_summary_sol} rendering the flow mostly potential. This situation is realized for charge transport sufficiently far away from charge neutrality~\cite{AKS,Lucas}.

Below we obtain the transport coefficients that are valid in the entire crossover region between these two regimes. We work to leading order accuracy in the parameter $\varepsilon$ in Eq.~\eqref{eq:locality_condition}.   In this approximation, we may neglect the wave number dependence of  $\lambda_q$ in Eq.~\eqref{eq:lambda_q_def}, setting  $\lambda_q \to \lambda_0$
in the subsequent expressions of the previous section.

In particular, the  matrix $\hat{K}_q$ in Eq. \eqref{eq:K_hat} can be reduced to a more compact expression:
\begin{eqnarray}\label{eq:zeta_hat_long_wavelengths}
  &&\hat{K}_0 =    \frac{1}{  \left( i \hat{\tau}_y \vec{x}_0 \right)^{\mathbb{T}}\hat{\Upsilon}_0  (i \hat{\tau}_y \vec{x}_0) }  \, (i \hat{\tau}_y \vec{x}_0) \otimes \left( i \hat{\tau}_y \vec{x}_0 \right)^{\mathbb{T}} \nonumber \\
  && = \frac{1}{\frac{\sigma_0}{e^2} s_0^2 - 2\frac{\gamma_0}{T}  n_0 s_0 + \frac{\kappa_0}{T} n_0^2}
  \left(
    \begin{array}{cc}
      s_0^2 & - n_0 s_0 \\
      - n_0 s_0 & n_0^2 \\
    \end{array}
  \right).
\end{eqnarray}
As a result, Eq.~\eqref{eq:friction_coefficient} for the friction coefficient simplifies to
\begin{eqnarray}\label{eq:friction_coefficient_long}
  k = \frac{\left\langle   \left(  s_0 \delta n - n_0 \delta s \right)^2  \right\rangle }{2 \left( \frac{\sigma_0}{e^2} s_0^2 - 2\frac{\gamma_0}{T}  n_0 s_0 + \frac{\kappa_0}{T} n_0^2 \right)}.
\end{eqnarray}
The friction coefficient $k$ here is positive definite because the denominator in Eq.~\eqref{eq:friction_coefficient_long} is positive. The latter statement follows from the fact that the matrix $\hat{\Upsilon}$ of kinetic coefficients of the liquid is positive definite. Furthermore, from the positivity of $k$ in Eq.~\eqref{eq:friction_coefficient_long}, it follows that  the friction coefficient in the general expression of Eq. \eqref{eq:friction_coefficient} is also positive definite~\cite{footnote}.

Next, we note that in the long wavelength approximation \eqref{eq:locality_condition} the column vector of renormalized densities $\vec{x}_{\rm{e}}$ in Eq.~\eqref{eq:x_eff} can  be expressed using the matrix $\hat{K}_0$ in Eq.~\eqref{eq:zeta_hat_long_wavelengths} in terms of the local fluctuations of densities, $ \delta \vec{x}$  and kinetic coefficients, $\delta \hat{\Upsilon} $, in the form
\begin{equation}\label{eq:x_e_long_wavelength}
  \vec{x}_{\rm{e}} = \vec{x}_0 - \frac{1}{2} \left\langle \delta \hat{\Upsilon} \hat{K}_{0} \delta \vec{x}  \right\rangle    - \frac{1}{2 \lambda_{0}} \left\langle  \delta \vec{x} \otimes \delta \vec{x}^{\mathbb{T}} \right\rangle   \hat{\Upsilon}^{-1}_0 \vec{x}_0.
\end{equation}

Let us now turn to the thermoelectric conductivity matrix. Note that since the friction coefficient $k$ is quadratic in the disorder amplitude, the first two terms in Eq.~\eqref{eq:Upsilon_eff_general} do not vanish in the limit of weak disorder. In contrast, the remaining terms in Eq. \eqref{eq:Upsilon_eff_general} are proportional to the variance of the disorder potential and are generally small in comparison to the first two. However, while most of these terms can be expressed in terms of local correlators, and are independent of disorder correlation radius $\xi$, the first term in the last line of Eq.~\eqref{eq:Upsilon_eff_general}, which is inversely proportional to the shear viscosity $\eta$, scales as $\xi^2$. Therefore, for systems where the correlation radius of disorder satisfies the condition Eq. \eqref{eq:locality_condition}, we may neglect all terms in Eq.~\eqref{eq:Upsilon_eff_general}  that are proportional to the disorder variance except for the first term in the last line.  Doing so, we obtain
\begin{eqnarray}\label{eq:Upsilon_eff_local}
 \hat{\Upsilon}_{\rm{e}} & = &\frac{1}{k}\big(\vec{x}_{\rm{e}} \otimes \vec{x}_{\rm{e}}^{\mathbb{T}}\big)  +   \hat{\Upsilon}_0   +    \int_{\bf{q}}
  \frac{ 1 }{ 2\eta q^2 }  \left( \vec{x}_{-\q} \otimes \vec{x}_\q^{\mathbb{T}}\right) .
\end{eqnarray}
Here $k$  and $\vec{x}_{\rm{e}} $ are given, respectively, by Eqs.~\eqref{eq:friction_coefficient_long}   and \eqref{eq:x_e_long_wavelength}, and within the accuracy of our approximation only terms of zeroth and second order in the fluctuations should be retained in $\big(\vec{x}_{\rm{e}} \otimes \vec{x}_{\rm{e}}^{\mathbb{T}}\big)$. The matrix in Eq. \eqref{eq:Upsilon_eff_local} is obviously positive definite.

As an application of Eq.~\eqref{eq:Upsilon_eff_local}, let us consider the  electrical resistivity $\rho$ of the system at weak long-range disorder.
It can be expressed in terms of  the $11$ matrix element of $\hat{\Upsilon}_{\rm{e}}$  in Eq.~\eqref{eq:Upsilon_eff_local} as $\rho = 1/ e^2 [\hat{\Upsilon}_{\rm{e}}]_{11} $. To leading accuracy in the disorder strength and in the long wavelength limit, this yields $\rho\approx k/(e^2n^2_0)$. Substituting here $k$ from Eq. \eqref{eq:friction_coefficient_long}, we get
\begin{equation}\label{eq:rho_nG}
\rho=\frac{1}{2e^2}\frac{T  \langle[\delta(s/n)]^2\rangle }{\kappa_0 - 2 \gamma_0 \frac{s_0}{n_0} + T \frac{\sigma_0}{e^2} \left( \frac{s_0}{n_0}\right)^2}.
\end{equation}
This expression generalizes the result of Ref.~\cite{AKS} in the long wavelength limit. Setting  $(\sigma_0, \gamma_0) =0$ in Eq.~\eqref{eq:rho_nG}, we reproduce the Galilean-invariant result \cite{AKS}  in the long wavelength limit.

The expressions for the effective friction coefficient, Eq.~\eqref{eq:friction_coefficient_long},  disorder-renormalized densities, Eq. \eqref{eq:x_e_long_wavelength}, and  thermoelectric conductivity matrix, Eq. \eqref{eq:Upsilon_eff_local},  are the main results of this section. They are applicable to systems with weak, $(\delta n , \delta s) \ll \mathrm{max} \{ |n_0|, s_0\}$, long-range disorder whose correlation length $\xi$ satisfies the condition Eq. \eqref{eq:locality_condition}.

Note that the first term in Eq.~\eqref{eq:Upsilon_eff_local} has the form of a projection operator on the column vector of  effective densities of particles and entropy in Eq.~\eqref{eq:x_e_long_wavelength}. This term may be identified with disorder-renormalized hydrodynamic transport mode. The corresponding conductivity is inversely proportional to the friction coefficient $k$ in Eq.~\eqref{eq:friction_coefficient_long}, and diverges at vanishing disorder.  The remaining terms in Eq.~\eqref{eq:Upsilon_eff_local} are associated with   disorder-renormalized transport mode relative to the liquid.  The first of these terms arises from the intrinsic transport relative to the liquid. The second term represents the contribution of convective vortical flow of particles and entropy induced on spatial scales of order of the correlation radius of disorder.  Although this contribution is proportional to the disorder strength, it grows as $\xi^2$ as the correlation radius increases. Therefore, the perturbative smallness is compensated in this term by the large parameter $1/\varepsilon$.

The conditions $(\delta n , \delta s) \ll \mathrm{max} \{ |n_0|, s_0\}$ and Eq. \eqref{eq:locality_condition} that define the applicability of Eqs. \eqref{eq:x_e_long_wavelength}--\eqref{eq:Upsilon_eff_local} ensure that the neglected nonlocal corrections to the hydrodynamic mode and to the relative mode separately, are relatively small. However, in a particular transport setup, both of these two modes may provide a contribution. In this case our approximation, which involves retaining the last two terms in Eq.~\eqref{eq:Upsilon_eff_local} while neglecting the nonlocal corrections to the first term, requires further justification. Let us consider the electrical conductivity of the system as an example. The nonlocal corrections to the main term in Eq.~\eqref{eq:Upsilon_eff_local} come from two sources: (i) corrections to the friction coefficient, and (ii) corrections to the effective density. Using  the form of the friction matrix in Eq. \eqref{eq:K_hat}, the nonlocal correction to the friction coefficient in Eq. \eqref{eq:friction_coefficient} can be  estimated as $\delta k\sim \langle\delta n^2\rangle[(\eta+\zeta)n^2_0]/(\xi\lambda_0)^2$. This yields the correction to the conductivity $\delta\sigma_k\sim n^2_0\delta k/k^2$.  The modification of conductivity due to nonlocal corrections to the effective density are of the same order of magnitude. On the other hand, the conductivity enhancement due to vortical viscous flow that is determined by the third term in Eq. \eqref{eq:Upsilon_eff_local} is estimated as $\delta\sigma_\eta\sim\xi^2\langle\delta n^2\rangle/\eta$. Thus the last term in Eq.~\eqref{eq:Upsilon_eff_local} exceeds the nonlocal corrections to the first term  provided the following condition is satisfied:
\begin{equation}\label{eq:applicability_condition_general}
  n_0^2 < \frac{\xi^2 k \lambda_0}{\sqrt{\eta (\eta + \zeta)}}.
\end{equation}
Note here that to the leading order $k\sim\langle\delta n^2\rangle$ per Eq. \eqref{eq:friction_coefficient_long}. We will consider the implications of the above condition in greater detail in the next section where we analyze electron transport in graphene near charge neutrality.


\section{Graphene near charge neutrality}\label{sec:graphene}

We now apply the principal results of Sec.~\ref{sec:long_range} to study the thermoelectric properties of graphene  near the charge neutrality point. Our perturbative approach  assumes that the variations of density are small, $\langle \delta n^2 \rangle \ll s_0^2+ n_0^2$. This condition can be satisfied in boron nitride encapsulated graphene devices. Furthermore, it is easy to see that for graphene near charge neutrality the condition Eq. \eqref{eq:locality_condition} is satisfied in the regime of applicability of the hydrodynamic approximation, $v /(T \xi) \ll 1$. Therefore, transport properties of graphene devices can be investigated using the long-range approximation of Sec.~\ref{sec:long_range}. Indeed, near charge neutrality, we can estimate $\eta q^2 \sim T^2/( v \xi)^2$, where $v$ is the band velocity in graphene, and $\xi$ is the correlation length of the disorder potential. Here we took $\eta \sim T^2/v^2$ \cite{Muller} and suppressed all logarithmic renormalizations in $\eta$ which are beyond the accuracy of these estimates.

One can further simplify the expressions of Sec.~\ref{sec:long_range}  by neglecting small terms of order $n_0/s_0$, which is valid in the regime near charge neutrality, $n_0 \ll s_0$.
In this approximation, the  friction coefficient, $k_0=k(n_0\to0)$,  in Eq.~\eqref{eq:friction_coefficient} can be expressed in terms of the intrinsic conductivity and the electron density variance as follows:
\begin{equation}\label{eq:k_graphene}
  k_0 = \frac{e^2}{\sigma_0}\frac{\langle \delta n^2\rangle}{2}.
\end{equation}
Furthermore, the expression Eq. \eqref{eq:x_e_long_wavelength} for the vector of renormalized densities simplifies to
\begin{equation}\label{eq:x_e_graphene}
  \vec{x}_{\rm{e}} = \left(
                \begin{array}{c}
                  n_0 \\
                  s_0 - \frac{e^2}{2 \sigma_0 T} \langle  \delta n \delta \gamma\rangle \\
                \end{array}
              \right).
\end{equation}

Substituting the expressions Eqs. \eqref{eq:k_graphene} and \eqref{eq:x_e_graphene} for the friction coefficient and disorder-renormalized densities of graphene into Eq. \eqref{eq:Upsilon_eff_local}, we  can express the thermoelectric conductivity matrix in the form
\begin{align}
\label{eq:Upsilon_eff_graphene}
& \hat{\Upsilon}_{\rm{e}}      \approx   \frac{2 \sigma_0}{e^2 \langle \delta n^2 \rangle}
  \left(
    \begin{array}{cc}
      n_0^2  & n_0 s_0 \\
      n_0 s_0 & s_0^2 \\
    \end{array}
  \right) \nonumber \\
  &
  - \frac{\langle \delta n \delta \gamma \rangle}{ T \langle\delta n^2 \rangle}
    \left(
    \begin{array}{cc}
      0  & n_0  \\
      n_0  & 2 s_0 \\
    \end{array}
  \right)    +
  \left(\begin{array}{cc}
   \sigma_0/e^2  + \chi & \gamma_0/T \\
   \gamma_0/T & \kappa_0/T  \\
    \end{array}\right),
\end{align}
where we introduced a dimensionless quantity:
\begin{equation}
\label{eq:chi_def}
\chi = \frac{1}{2\eta} \int \frac{d^2 q}{(2\pi)^2}  \frac{| n_\q|^2}{q^2} .
\end{equation}

\begin{figure}[t!]
  \centering
  \includegraphics[width=3.25in]{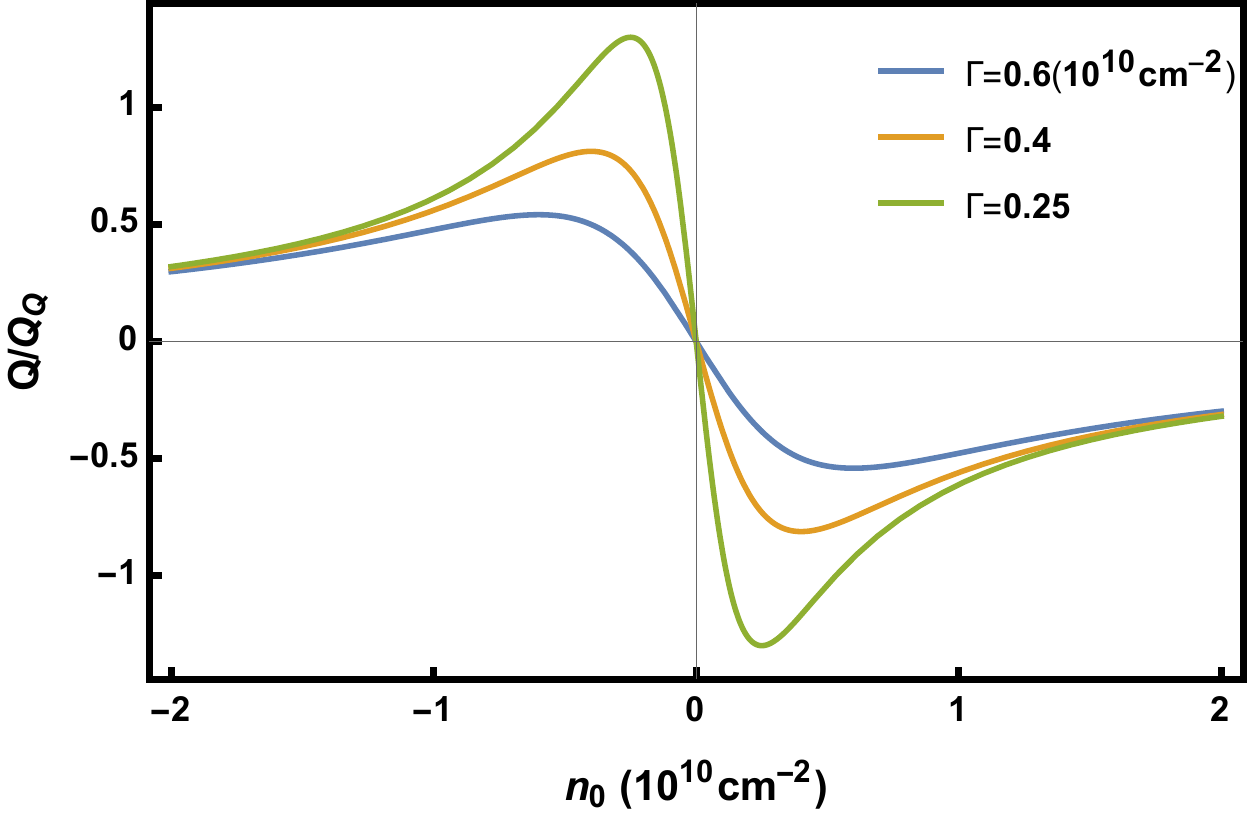}
  \includegraphics[width=3.25in]{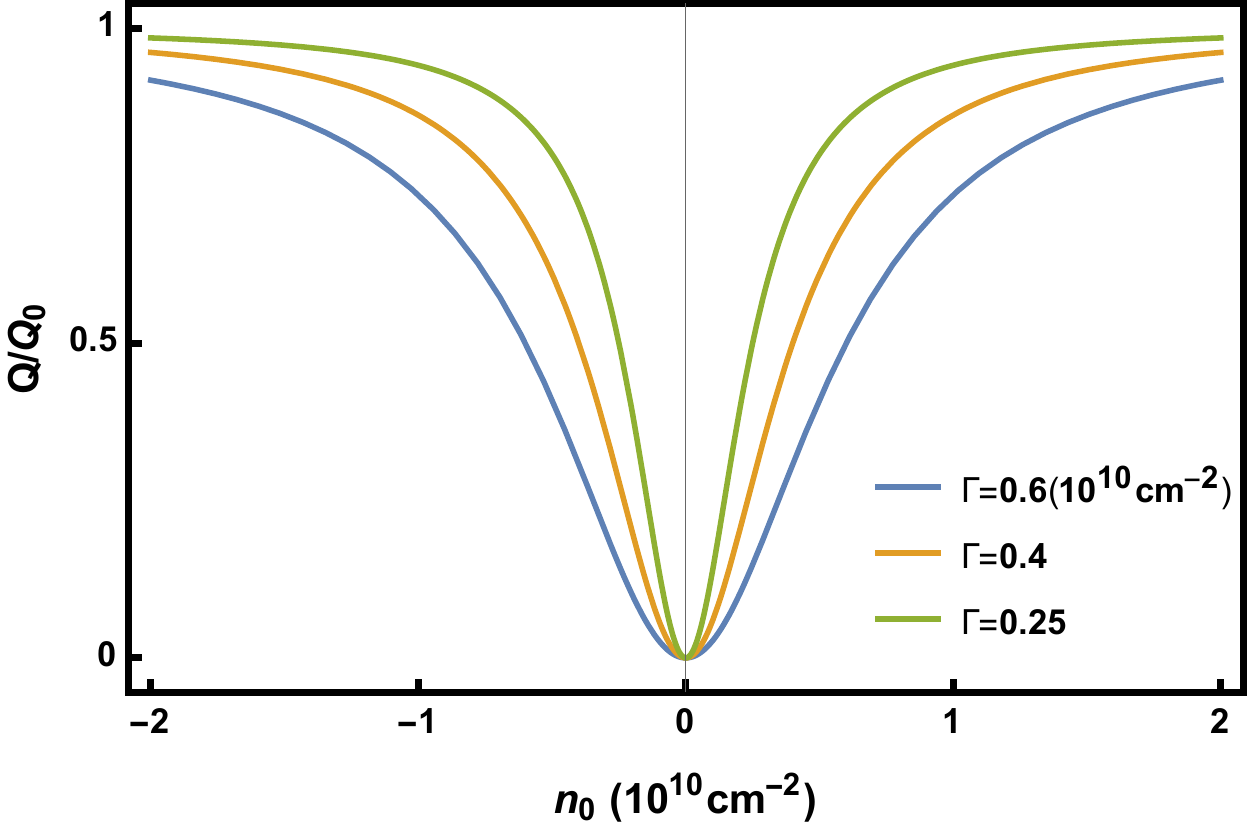}
  \caption{[Upper panel]: Density dependence of the Seebeck coefficient in Eq.~\eqref{eq:Seebeck_graphene} for several different values of the disorder strength as quantified by a parameter $\Gamma=\sqrt{\frac{1}{2}\langle\delta n^2\rangle(1+\frac{e^2}{\sigma_0}\chi)}$. Note that $\Gamma$ is strongly temperature dependent so the sensitivity to thermal fluctuations are implicit (see further discussion in Sec. \ref{sec:summary}). The scale of $Q$ on the plot is normalized to a ``quantum" unit of thermopower $Q_Q=\pi^2/3$ (in full dimensional units it is $\pi^2k^2_B/3e$). [Lower panel]: The same plot as above but with the different normalization of the overall scale to the prediction of an ideal hydrodynamic limit $Q_Q\to Q_{0}$. This plot emphasizes that as a function of density (or temperature implicit in $\Gamma$), thermopower is enhanced as compared to prediction of the Mott formula but stays below $Q_{0}$.}
  \label{Fig-Q}
\end{figure}

The electrical conductivity of the system is defined by the $11$ element of the matrix $\hat{\Upsilon}$  in Eq.~\eqref{eq:Upsilon_eff_graphene}, and is given by
\begin{equation}\label{eq:conductivity_greaphene_n}
  \sigma = \sigma_0 + e^2 \chi + \sigma_0 \frac{2 n_0^2}{\langle \delta n^2 \rangle }.
\end{equation}
Note that long-range correlated disorder \emph{enhances} the conductivity of the system at charge neutrality.

To obtain the thermoelectric properties of the systems, we re-express the linear relation Eq. \eqref{eq:Upsilon_eff_def} with the matrix coefficients Eq. \eqref{eq:Upsilon_eff_graphene} in a more familiar form~\cite{Abrikosov}:
\begin{subequations}\label{eq:Abrikosov}
  \begin{eqnarray}
 \langle  \mathbf{\mathcal{E}} \rangle &=&  \rho_{\mathrm{e}}\,   e\langle \mathbf{j} \rangle   + Q_{\mathrm{e}} \langle \bm{\nabla} T, \rangle\\
  T \langle \mathbf{j}_s \rangle  &=&  \Pi_{\mathrm{e}}  e\langle \mathbf{j} \rangle  -  \kappa_{\mathrm{e}}\langle \bm{\nabla} T \rangle.
\end{eqnarray}
\end{subequations}
Here $\rho_{\mathrm{e}}$ and $ \kappa_{\mathrm{e}}$ are the electrical resistivity and the thermal conductivity of the system, and $Q_{\mathrm{e}}$ and $\Pi_{\mathrm{e}}$ are, respectively, the Seebeck coefficient (thermopower) and the Peltier coefficient. The latter are related by the Onsager symmetry relation $\Pi_{\mathrm{e}}= Q_{\mathrm{e}} T$.

For the Peltier and  Seebeck coefficients, we obtain to leading order accuracy in inhomogeneity:
\begin{equation}\label{eq:Seebeck_graphene}
  Q_{\mathrm{e}} = \frac{\Pi_{\mathrm{e}}}{T}= \frac{1}{e} \, \frac{2 n_0 s_0}{\langle \delta n^2 \rangle}\frac{1   }{ 1  + \frac{e^2}{\sigma_0} \chi + \frac{2n^2_0}{\langle \delta n^2 \rangle} }.
\end{equation}
The Seebeck coefficient $Q_{\mathrm{e}}$ is given by the entropy per unit charge that is transported by the current. At relatively large doping, $n_0^2 \gg \langle \delta n^2 \rangle$,  it approaches the value in the pristine electron liquid, $  Q_0=\frac{1}{e}\, \frac{s_0}{ n_0} $ but is always reduced from it. This reduction is especially strong,  near charge neutrality. The doping dependence of the Seebeck coefficient is illustrated in Fig.~\ref{Fig-Q}.

\begin{figure}[t!]
  \centering
  \includegraphics[width=3.25in]{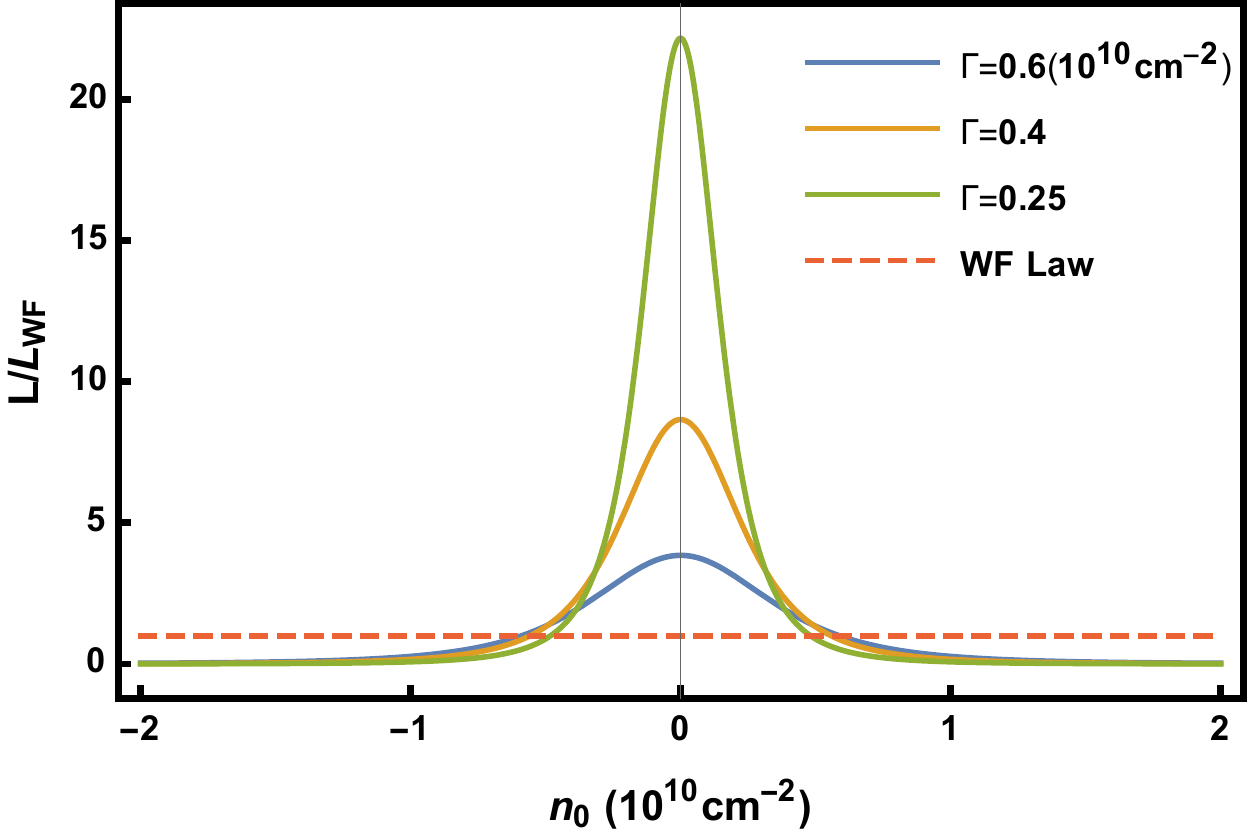}
  \caption{Density dependence of the Lorentz ratio in Eq.~\eqref{eq:Lorentz_ratio_gr} (scaled with the Wiedemann-Franz value $L_{\text{WF}}=\pi^2/3e^2$)  for three different values of the disorder-induced density fluctuations as quantified by the parameter $\Gamma=\sqrt{\frac{1}{2}\langle\delta n^2\rangle(1+\frac{e^2}{\sigma_0} \chi )}$.}
  \label{Fig-L}
\end{figure}

Finally, for the thermal conductivity, we obtain
\begin{equation}
\label{eq:thermal_conductivity_gr}
\kappa_{\mathrm{e}} = T  \,  \frac{2 s_0^2}{\langle \delta n^2 \rangle}
\frac{\frac{\sigma_0}{e^2}  +  \chi }{ 1  + \frac{e^2}{\sigma_0} \chi   + \frac{2n^2_0}{\langle \delta n^2 \rangle}}.
\end{equation}
This yields the Lorentz ratio $ L_{\mathrm{e}}=\kappa_{\mathrm{e}} \rho_{\mathrm{e}} /T$ in the form
\begin{equation}\label{eq:Lorentz_ratio_gr}
  L_{\mathrm{e}}= \frac{2 s_0^2}{e^2\langle \delta n^2 \rangle}
  \frac{ 1 +  \frac{e^2}{\sigma_0}\chi }{ \left[1  + \frac{e^2}{\sigma_0} \chi   + \frac{2n^2_0}{\langle \delta n^2 \rangle}\right]^2}.
\end{equation}
The density dependence of the Lorentz ratio in Eq.~\eqref{eq:Lorentz_ratio_gr} is illustrated in Fig.~\ref{Fig-L}.

Equations \eqref{eq:Upsilon_eff_graphene}, \eqref{eq:conductivity_greaphene_n}, and \eqref{eq:Seebeck_graphene}--\eqref{eq:Lorentz_ratio_gr} for the transport coefficients of graphene near charge neutrality represent the main results of this section. Note that the disorder affects the transport coefficients of the system in two qualitatively different ways: (i) via the friction coefficient $k$ in Eq.~\eqref{eq:k_graphene}, and (ii) via the parameter $\chi$ in Eq.~\eqref{eq:chi_def}.

(i) The first part of the dependence can be obtained  by introducing a friction force in the hydrodynamic equations \cite{Hartnoll,Sachdev,Aleiner,XL,Xie-Foster,Principi-2DM,Danz}. This approach produces viscosity-independent  transport coefficients that correspond to setting $\chi\to 0$ in our expressions. In particular, it yields the conductivity at charge neutrality that is unaffected by disorder and equal to the intrinsic conductivity of the electron liquid~\cite{Hartnoll,Sachdev,Aleiner}. The reason is that within such an approach electron transport at charge neutrality is decoupled from the hydrodynamic flow.

(ii) The dependence of the transport coefficients on the shear viscosity of the electron liquid is described by the parameter $\chi$ in Eq.~\eqref{eq:chi_def}. Although $\chi$ is proportional to the disorder strength, it scales with the correlation radius as $\xi^2$, and thus is inversely proportional to the small parameter $\varepsilon$ in Eq.~\eqref{eq:locality_condition}. This shows that the effect of the long-range disorder on hydrodynamic electron transport near charge neutrality is extremely nonlocal. This is qualitatively different from the situation away from charge neutrality, where the resistivity becomes independent of the correlation radius for long-range disorder~\cite{AKS,Lucas}. The dependence of the transport coefficients on $\chi$ arises from the vortical component of the inhomogeneous flow described by Eq.~\eqref{eq:u_transverse_summary_sol}. This vortical flow gives a positive contribution to the conductivity of the system, an enhancement factor $e^2\chi$ in Eq.~\eqref{eq:conductivity_greaphene_n}, that is inversely proportional to the shear viscosity $\eta$.

Note that accounting for the dependence of $\lambda_q$ on the wave number $q$ in our general expressions would also produce viscosity-dependent corrections to the resistivity of the system. These corrections arise from the potential component of the flow, and scale as $(\eta + \zeta)/\xi^2$, cf. Refs. \cite{AKS,Lucas,LXA}. Their contribution to the electrical conductivity is proportional to $n_0^2$ and becomes essential away from charge neutrality. The condition that these corrections are small in comparison to the second term in Eq.~\eqref{eq:conductivity_greaphene_n} is expressed by Eq.~\eqref{eq:applicability_condition_general}. Using Eq.~\eqref{eq:k_graphene} and the estimates at the beginning of this section, it is easy to see that the applicability condition Eq. \eqref{eq:applicability_condition_general} reduces to
\begin{equation}\label{eq:applicability_condition_graphene}
  |n_0| \lesssim \frac{\xi}{l_T} \sqrt{ \langle\delta n^2\rangle},\quad l_T=\frac{v}{T}.
\end{equation}
This condition is satisfied in a parametrically wide range of electron doping near charge neutrality, in which the viscous contribution to  resistivity  is dominated by vortical flow at spatial scales comparable to $\xi$. This flow produces a positive contribution to the electrical conductivity of the system, which is proportional to the square of the disorder correlation radius. In contrast, sufficiently far from charge neutrality, the viscous contribution to the conductivity arises predominantly from the nonlocal corrections to Eq.~\eqref{eq:Upsilon_eff_graphene} due to the momentum dependence of $\lambda_q$ in the general expression Eq. \eqref{eq:Upsilon_eff_general}. They correspond to  the longitudinal component of the hydrodynamic flow,  and give a positive contribution to the resistivity that is proportional to the combination $(\eta + \zeta)/\xi^2$ \cite{AKS,Lucas,LXA}.

It is important to note that the dimensionless parameter $\chi$ in Eq.~\eqref{eq:chi_def}, which can be estimated near charge neutrality as $\chi \sim (\xi l_T)^2\langle \delta n^2\rangle$, may become large even at weak, $\langle \delta n^2\rangle \ll s_0^2$, but sufficiently long-range disorder, $\xi\gg l_T$.  This imposes additional constraints on the applicability of our results. They may be obtained by considering an ``isotropic" extension of the toy model from the qualitative discussion in  Sec. \ref{sec:intro} with a checkerboard pattern of inhomogeneous density, $ n (\mathbf{r}) =\delta n_0 [\cos ( y /\xi) + \cos ( x /\xi) ]$. In this case, in addition to the viscous force, an inhomogeneous flow will cause the friction force, whose magnitude can be estimated from Eq.~\eqref{eq:k_graphene} as $k u_0 \sim u_0 (e^2/\sigma_0) \delta n_0^2$. Our results obtained in the long wavelength approximation of Sec.~\ref{sec:long_range} apply as long as the friction force is smaller than the viscous force, $e^2 \xi^2 \langle \delta n^2 \rangle /\eta \lesssim \sigma_0 $. This imposes the following constraint:
\begin{equation}\label{eq:chi_constraint}
   \chi \lesssim \sigma_0/e^2
\end{equation}
on the applicability of our results. This implies that our results remain valid as long as the enhancement of the conductivity at charge neutrality does not exceed the intrinsic value in Eq.~\eqref{eq:conductivity_greaphene_n}. The constraint Eq. \eqref{eq:chi_constraint} on the applicability of our perturbative results can also be verified by considering higher order terms in the perturbation theory.


\section{Summary}\label{sec:summary}

In this paper, we presented a hydrodynamic theory of electron transport in the conductors with a long-range disorder potential.  It generalizes the approach of Ref. \cite{AKS} to conductors in which the underlying electron liquid lacks Galilean invariance.

For weak disorder, the matrix of kinetic coefficients of the system defined by Eq. \eqref{eq:Upsilon_eff_def} can be expressed in terms of the position-dependent densities of particles and entropy Eq. \eqref{eq:x_X_def} and intrinsic kinetic coefficients Eq. \eqref{eq:gamma_def} of the liquid in the form of Eq. \eqref{eq:Upsilon_eff_general}.

For a long-range disorder potential, whose correlation radius satisfies the condition Eq.~\eqref{eq:locality_condition}, the transport coefficients may be expressed in a simplified form Eq. \eqref{eq:Upsilon_eff_local} in terms of the friction coefficient Eq.~\eqref{eq:friction_coefficient_long}, disorder-renormalized particle and entropy densities, and an additional viscosity dependent parameter $\chi$ defined in Eq.~\eqref{eq:chi_def}. The dependence of kinetic coefficients on $\chi$ that we find represents the principal difference of our results from those of the previous treatments~\cite{Hartnoll,Sachdev,Aleiner,XL,Xie-Foster,Principi-2DM,Danz}. The dependence of the transport coefficients on $\chi$ arises from the vortical component of the hydrodynamic flow. Importantly, this dependence has an extremely nonlocal character; the parameter  $\chi$ in Eq.~\eqref{eq:chi_def} is inversely proportional to the square of the disorder correlation radius. The results of the previous treatments can be obtained by setting $\chi\to 0$ in our expressions.

For graphene devices subjected to long-range disorder, the transport coefficients are described by Eqs.~\eqref{eq:conductivity_greaphene_n}--\eqref{eq:Lorentz_ratio_gr}. Remarkably, the conductivity at charge neutrality, $n_0\to0$ in Eq.~\eqref{eq:conductivity_greaphene_n}, is enhanced in comparison to the intrinsic conductivity of the electron liquid. This is in contrast to previous results \cite{Lucas,Hartnoll,Sachdev,Aleiner}, which predict the conductivity at charge neutrality to be unaffected by disorder and equal to the intrinsic conductivity of the electron liquid. The conductivity enhancement arises from the convective charge transport by the vortical component of the hydrodynamic flow.

\subsection*{Acknowledgments}

We gratefully acknowledge illuminating discussions with I. Aleiner, K. C. Fong, P. Kim, S. Kivelson, L. Levitov, K. A. Matveev, A. Principi, and B. Spivak of various physical phenomena relevant to this work. This work was supported by the U.S. Department of Energy Office of Science, Basic Energy Sciences under Award No. DE-FG02-07ER46452 and by the National Science Foundation Grant MRSEC DMR-1719797 (A. V. A). S. L. and A. L. acknowledge support by the National Science Foundation CAREER Grant No. DMR-1653661 and in part by the Ray MacDonald Endowment Award at the UW-Madison. This work was performed in part at Aspen Center for Physics, which is supported by National Science Foundation Grant PHY-1607611.

\appendix

\section{Entropy production rate}
\label{sec:entropy_production}
In this Appendix, we derive the expression for the local entropy production rate in Eq.~\eqref{eq:s_dot_local_currents}. Assuming local thermal equilibrium, we can characterize the state of the liquid by the densities of conserved quantities: particle number $n$, energy $\epsilon$, and  momentum $\mathbf{p}$. The entropy density $s$ is a function of the conserved quantities, and its differential is given  by the thermodynamic relation
\begin{equation}\label{eq:d_s}
 ds =  \frac{d\epsilon}{T}  - \frac{( \mu + e\phi + U) d n}{T}  -  \frac{\mathbf{u}\cdot d \mathbf{p}}{T},
\end{equation}
where $U$ is the external potential, and the local equilibrium parameters $T$, $\mu$, and $\mathbf{u}$ are functions of $n$, $s$, and $\mathbf{p}$ which depend on the band structure. The electric potential $\phi$ is determined by the  density of electrons and external charges:
\begin{equation}\label{eq:phi_n}
  e\phi (\r) = \int d \r' \frac{e^2 [n(\r') + n_{\rm ext}(\r')]}{|\r-\r'|} .
\end{equation}
Using the thermodynamic relation Eq.~(\ref{eq:d_s}), and the evolution equations \eqref{eq:continuity_n} and \eqref{eq:continuity_p} for the conserved quantities, we get
\begin{eqnarray}\label{eq:s_dot}
  \partial_t s &=& - \frac{1}{T}\Big[\boldsymbol{\nabla}\cdot \mathbf{j}_\epsilon  - (\mu + e\phi + U)\,\boldsymbol{\nabla}\cdot \mathbf{j}  \nonumber \\
  && - u_i \partial_j \Pi_{ij}  - n \mathbf{u} \cdot \boldsymbol{\nabla} (e\phi + U)\Big] .
\end{eqnarray}
Using Eqs.~\eqref{eq:j_S_def} and \eqref{eq:constitutive_Pi} ,we can rewrite the evolution equation \eqref{eq:s_dot} in the form of Eq.~\eqref{eq:s_dot_source} where the expression for the local entropy production rate $\varsigma$
 is given by
\begin{eqnarray}
\label{eq:s_dot_local}
\varsigma & =& \mathbf{j}_\epsilon \cdot \boldsymbol{\nabla} \frac{1}{T} - \mathbf{j} \cdot \boldsymbol{\nabla} \left( \frac{\mu + e\phi + U}{T} \right)  \nonumber \\
&& + \frac{\mathbf{u} \cdot \boldsymbol{\nabla} P + n \mathbf{u} \cdot \boldsymbol{\nabla} (e\phi + U) - u_i \partial_j \sigma'_{ij}}{T} .
\end{eqnarray}
As the next step, using the thermodynamic identity $\boldsymbol{\nabla} P= n \boldsymbol{\nabla} \mu + s \boldsymbol{\nabla} T$ we can rewrite Eq.~\eqref{eq:s_dot_local} in the form
\begin{eqnarray}
\label{eq:s_dot_local_1}
\varsigma& =& - \frac{1}{T} \Big[\left(\mathbf{j}_s - s \mathbf{u}\right)\cdot \boldsymbol{\nabla} T + \left( \mathbf{j} - n\mathbf{u}\right) \cdot \boldsymbol{\nabla} \left(\mu + e\phi + U \right) \Big] \nonumber \\
&& - \frac{ u_i \partial_j \sigma'_{ij}}{T} .
\end{eqnarray}
Substituting the definitions Eqs. \eqref{eq:constitutive_j} and \eqref{eq:j_S'_def} of the dissipative particle and entropy fluxes we obtain the local entropy production rate in the form of Eq.~\eqref{eq:s_dot_local_currents}.

\section{Derivation of macroscopic transport coefficients from entropy production}
\label{sec:resistivity_matrix}

The macroscopic thermoelectric conductivity matrix defined in Eq.~\eqref{eq:Upsilon_eff_def} can be obtained by evaluating the entropy production rate in the system
\begin{eqnarray}
\label{eq:S_dot_total}
T \dot{S}&=& \int d \r \left[    \vec{\bf X}^{\mathbb{T}} \hat{\Upsilon}  \vec{\bf X} + \sigma'_{ij} \partial_i u_j \right],
\end{eqnarray}
and expressing it in terms of the macroscopic electric field and temperature gradient,
 $\vec{\mathbf{X}}_0$,  in the form
\begin{equation}
\frac{T\dot{S}}{A}= \vec{\mathbf{X}}^{\mathbb{T}}_{0}\hat{\Upsilon}_{\rm{e}}\vec{\mathbf{X}}_0.
\end{equation}
 This procedure is equivalent to the consideration in the main text.

To the second order in perturbation in disorder this yields
\begin{align}
&\vec{\mathbf{X}}^{\mathbb{T}}_{0}\hat{\Upsilon}_{\rm{e}}\vec{\mathbf{X}}_0=\vec{\mathbf{X}}^{\mathbb{T}}_{0}\hat{\Upsilon}_0\vec{\mathbf{X}}_0+\int_{\bf{q}}\left[
\vec{\mathbf{X}}^{\mathbb{T}}_{-\bf{q}}\hat{\Upsilon}_0\vec{\mathbf{X}}_{\bf{q}}+2\vec{\mathbf{X}}^{\mathbb{T}}_{0}\hat{\Upsilon}_{-\bf{q}}\vec{\mathbf{X}}_{\bf{q}}
\right. \nonumber\\
&\left.+\eta q^2 (\mathbf{u}^t_{\bf{q}}\cdot\mathbf{u}^t_{-\bf{q}})+(\eta+\zeta) q^2 (\mathbf{u}^l_{\bf{q}}\cdot\mathbf{u}^l_{-\bf{q}})\right] .
\label{eq:S_dot_approx}
\end{align}
Substituting Eqs. \eqref{eq:solution_summary} into this equation, and using Eqs. \eqref{eq:x_eff}--\eqref{eq:K_hat} we reproduce the result of Eq. \eqref{eq:Upsilon_eff_general} for $\hat{\Upsilon}_{\rm{e}}$.

Finally, to establish a direct connection to the macroscopic thermoelectric resistivity matrix, one can rewrite entropy production rate in a different way, $T\dot{S}/A =  \langle \vec{\bf J}\rangle^{\mathbb{T}} \hat{\varrho}\langle \vec{\bf J}\rangle$,
Indeed, one should simply recall that the matrix  $\hat{\varrho}$ defines a linear relation, $ \langle \vec{\mathbf{X}} (\r)\rangle =\hat{\varrho} \langle \vec{\bf J}\rangle $, between the average currents $\langle \mathbf{j} \rangle, \langle \mathbf{j}_s \rangle$ and the average forces $e\langle\boldsymbol{\mathcal{E}}\rangle,-\langle\boldsymbol{\nabla}T\rangle$. This matrix must satisfy the Onsager symmetry principle as can be readily verified based on Eq. \eqref{eq:Upsilon_eff_general}.


\begin{thebibliography}{99}
\bibitem{Note}
The electron density is counted modulo fully occupied bands, since those do not contribute to electron transport.
\bibitem{footnote}
Indeed, the long wavelength approximation underestimates the denominator in the second term in the first line of Eq.~\eqref{eq:friction_coefficient}. Thus, the long wavelength approximation Eq. \eqref{eq:friction_coefficient_long} provides a positive lower bound for the friction coefficient in Ref. \eqref{eq:friction_coefficient}.
\end{thebibliography}
\end{document}